\documentclass[twocolumn,letter]{jpsj3}
\usepackage{graphicx}
\usepackage{dcolumn}
\usepackage{bm}
\usepackage{ulem}
\usepackage{braket}
\usepackage{amsmath}
\usepackage{color}

\title{Spin Conductivity Based on Magnetic Toroidal Quadrupole Hidden in Antiferromagnets}

\author{Satoru Hayami$^1$ and Megumi Yatsushiro$^{1,2}$}
\inst{
$^1$Department of Applied Physics, the University of Tokyo, Tokyo 113-8656, Japan \\
$^2$Department of Physics, Hokkaido University, Sapporo 060-0810, Japan 
}

\abst{
We report our theoretical results on spin conductivity in antiferromagnets by focusing on the role of the magnetic toroidal quadrupole (MTQ) in electron systems. 
The MTQ is characterized as a time-reversal-odd rank-2 polar tensor degree of freedom in electrons, which is distinct from conventional rank-1 magnetic and magnetic toroidal dipoles. 
Based on a microscopic $sd$ model analysis for a tetragonal system under both collinear and noncollinear antiferromagnetic orderings, we clarify that the MTQ becomes a source of an extrinsic spin conductivity even with neither a uniform magnetization nor spin-orbit coupling.  
We also list all the magnetic point groups to accommodate the MTQs as a primary order parameter as well as the candidate antiferromagnetic materials. 
}

\begin{document}
\maketitle

A magnetic toroidal (MT) moment, which corresponds to a time-reversal-odd polar tensor, is one of the fundamental moments as well as electric and magnetic moments~\cite{Spaldin_0953-8984-20-43-434203,kopaev2009toroidal,cheong2018broken}. 
Especially, the dipole component of the MT moment, i.e., the MT dipole (MTD), has been extensively studied in both theory and experiment, since it becomes a source of parity-violating physical phenomena in magnetic materials, such as the linear magnetoelectric effect~\cite{popov1999magnetic,arima2005resonant,van2007observation,zimmermann2014ferroic,Toledano_PhysRevB.92.094431}, nonreciprocal directional dichroism~\cite{Sawada_PhysRevLett.95.237402,Kezsmarki_PhysRevLett.106.057403,Miyahara_JPSJ.81.023712,Miyahara_PhysRevB.89.195145,Bordacs_PhysRevB.92.214441,Sato_PhysRevLett.124.217402}, nonlinear magnon spin Nernst effect~\cite{Kondo_PhysRevResearch.4.013186}, and nonreciprocal magnon excitations~\cite{Iguchi_PhysRevB.92.184419,Hayami_doi:10.7566/JPSJ.85.053705,Gitgeatpong_PhysRevLett.119.047201,sato2019nonreciprocal,Matsumoto_PhysRevB.101.224419,Matsumoto_PhysRevB.104.134420,Hayami_PhysRevB.105.014404}.
Although such MTD-related phenomena were originally investigated in magnetic insulators in the field of multiferroics, recent studies have clarified that the emergence of the MTD in magnetic metals results in similar multiferroic phenomena~\cite{Mentink1994,Yanase_JPSJ.83.014703,Hayami_PhysRevB.90.024432,Hayami_doi:10.7566/JPSJ.84.064717,thole2018magnetoelectric,Gao_PhysRevB.97.134423,Watanabe_PhysRevB.98.220412,Shitade_PhysRevB.98.020407}, nonreciprocal transport~\cite{yatsushiro2021microscopic,Suzuki_PhysRevB.105.075201,Hayami2022nonlinear}, spin-orbital-momentum locking~\cite{Hayami_PhysRevB.104.045117}, and nonlinear spin Hall effect~\cite{Hayami2022spinHall}, which extends the scope of MTD-related materials~\cite{saito2018evidence,shinozaki2020magnetoelectric,Yanagisawa_PhysRevLett.126.157201}.

\begin{figure}[t!]
\begin{center}
\includegraphics[width=1.0 \hsize ]{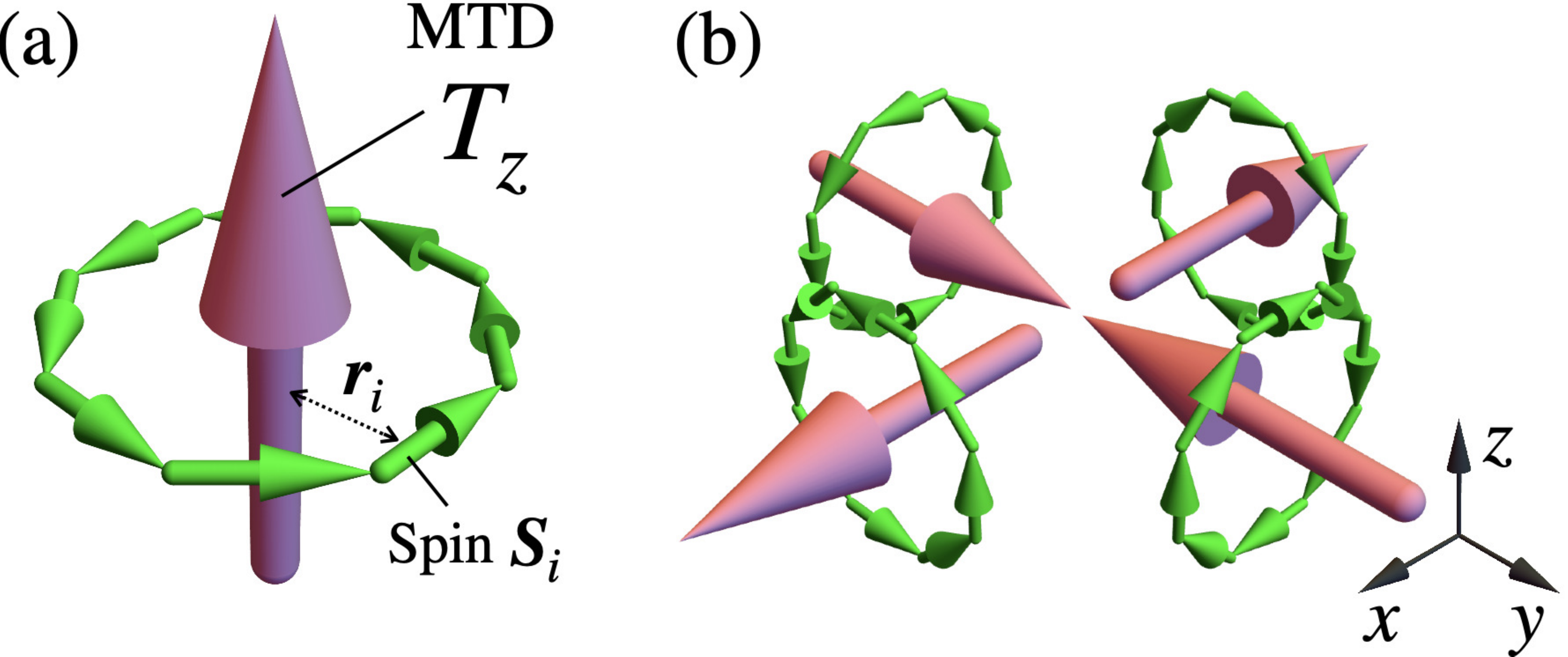} 
\caption{
\label{fig: fig1}
(Color online)
(a,b) Schematic pictures of the magnetic toroidal dipole (MTD) $T_z$ (a) and quadrupole (MTQ) $T_v$ (b), where the green and pink arrows represent spin and MTD moments, respectively. 
}
\end{center}
\end{figure}

The MTD has often been described by the vector product of the position vector $\bm{r}_i$ and the classical spin $\bm{S}_i$ at site $i$, whose expression is given by~\cite{dubovik1975multipole,dubovik1990toroid,Spaldin_0953-8984-20-43-434203,kopaev2009toroidal} 
\begin{align}
\label{eq:MTD}
\bm{T}=\frac{g \mu_{\rm B}}{2}\sum_i \bm{r}_i \times \bm{S}_i, 
\end{align}
where $g$ and $\mu_{\rm B}$ represent the $g$ factor and the Bohr magneton, respectively. 
Hereafter, we omit $g$ and $\mu_{\rm B}$ in the expression.
From Eq.~(\ref{eq:MTD}), the MTD emerges under a vortex spin configuration, as shown in Fig.~\ref{fig: fig1}(a), whose spatial inversion ($\mathcal{P}$) and time-reversal ($\mathcal{T}$) parities are odd owing to $\mathcal{P}\bm{r}_i= -\bm{r}_i$ and $\mathcal{T}\bm{S}_i= -\bm{S}_i$; the MTD is distinct from the magnetic dipole characterizing a time-reversal-odd axial vector quantity like spin.  
The MTD manifests itself in various descriptions based on the quantum mechanical-operator expressions~\cite{hayami2018microscopic,kusunose2020complete}: the orbital hybridization~\cite{yatsushiro2019atomic,watanabe2019charge} and bond current~\cite{Hayami_PhysRevLett.122.147602,Hayami_PhysRevB.101.220403,Hayami_PhysRevB.102.144441}. 

The concept of the MTD moment is extended to higher-rank MT moments, which are referred to as MT multipoles~\cite{dubovik1975multipole,dubovik1990toroid,nanz2016toroidal,hayami2018microscopic} or hyper-toroidal moments~\cite{planes2014recent}. 
Such higher-rank MT multipoles are described by a nonuniform spatial distribution of the MTD. 
For example, the expressions of the higher-rank MT multipoles for a magnetic cluster with $\bm{S}_i$ are given by using the spherical harmonics $Y_{lm}(\hat{\bm{r}})$ as~\cite{Suzuki_PhysRevB.95.094406,suzuki2018first,Suzuki_PhysRevB.99.174407,Huebsch_PhysRevX.11.011031} 
\begin{align}
\label{eq: MT multipole}
T_{lm}=c_l \sum_{i } (\bm{r}_i\times \bm{S}_i)\cdot \bm{\nabla}_i \left[ r^l_i Y_{lm}^* (\hat{\bm r}_i) \right], 
\end{align}
where $l$ and $m$ represent the azimuthal quantum number and magnetic quantum number ($-l \leq m\leq l$), and 
$c_l$ is the numerical coefficient. 
Since the spatial parity of $Y_{lm}(\hat{\bm{r}}_i)$ is given by $(-1)^{l}$, that of $T_{lm}$ depends on the rank; the even(odd)-rank MT multipole is invariant (variant) under the $\mathcal{P}$ operation. 
Thus, physical properties under even-rank MT multipoles are qualitatively different from those under odd-rank MT multipoles like MTD. 
Nevertheless, an even-rank MT system has been less studied compared to an odd-rank one, since its characteristic features have been masked owing to the absence of uniform vector quantity.

In this Letter, we investigate the nature of the even-rank MT multipoles in the antiferromagnetic (AFM) systems in order to explore the possibility of exhibiting intriguing physical phenomena even without the uniform magnetic dipole (axial-vector quantity) and MTD (polar-vector quantity). 
By focusing on the $l=2$ component of the MT multipole, i.e., the MT quadrupole (MTQ), in AFMs, we find that the emergence of the MTQ causes spin conductive phenomena. 
The mechanism does not rely on atomic spin-orbit coupling (SOC). 
This is qualitatively different from that in the noncentrosymmetric nonmagnetic systems, where the antisymmetric SOC plays an important role. 
Although the present mechanism is closely related to the previous findings in the SOC-free AFMs with the spin-split band structure~\cite{naka2019spin,hayami2019momentum, Naka_PhysRevB.103.125114,shao2021spin,seo2021antiferromagnetic, Gonzales_PhysRevLett.126.127701,  Gurung_PhysRevMaterials.5.124411}, we show that the nonzero spin conductivity survives even without the spin-split band structure.
We demonstrate it by exemplifying both the collinear and noncollinear AFM orderings in the tetragonal system. 
Moreover, we list all the magnetic point groups (MPGs) with the MTQs but without the magnetic dipole in addition to the candidate materials. 
Our results open up a new direction of AFMs as a spin current generator based on the MTQ, which stimulates further exploration of the functional materials applicable to spintronics. 

Let us start by showing the cluster-multipole expression of the MTQ in AFMs, which is obtained as the $l=2$ component in Eq.~(\ref{eq: MT multipole}): 
\begin{align}
\label{eq: Tu}
T_u &= \sum_i (2z_i t_i^z - x_i t_i^x -  y_i t_i^y), \\
\label{eq: Tv}
T_v &= \sqrt{3}\sum_i (x_i t_i^x -  y_i t_i^y),\\
\label{eq: Tyz}
(T_{yz}, T_{zx}, T_{xy}) &=[\sqrt{3}\sum_i (y_i t_i^z + z_i t_i^y), {\rm (cyclic)}], 
\end{align}
where $\bm{t}_i=\bm{r}_i \times \bm{S}_i$. 
The MTQ is described by the spatial distribution of the local MTD $\bm{t}_i$, as schematically plotted in the case of $T_v$ [Eq.~(\ref{eq: Tv})] in Fig.~\ref{fig: fig1}(b). 
All the MTQs have even $\mathcal{P}$ parity but odd $\mathcal{T}$ parity. 
Although such a transformation regarding $\mathcal{P}$ and $\mathcal{T}$ is common to that of the magnetic dipole (uniform magnetization), the transformation regarding other point group operations, such as the mirror and rotational operations, is different owing to the different rank of multipoles~\cite{Hayami_PhysRevB.98.165110, Yatsushiro_PhysRevB.104.054412}. 
In terms of the representation theory, the MTQs can belong to the different irreducible representations from the magnetic dipoles under an MPG.  
We find 22 MPGs with the finite MTQ but without the magnetic dipole, as discussed below (see Table~\ref{table:mp}), where pure MTQ-related physical phenomena are expected.

\begin{figure}[t!]
\begin{center}
\includegraphics[width=1.0 \hsize ]{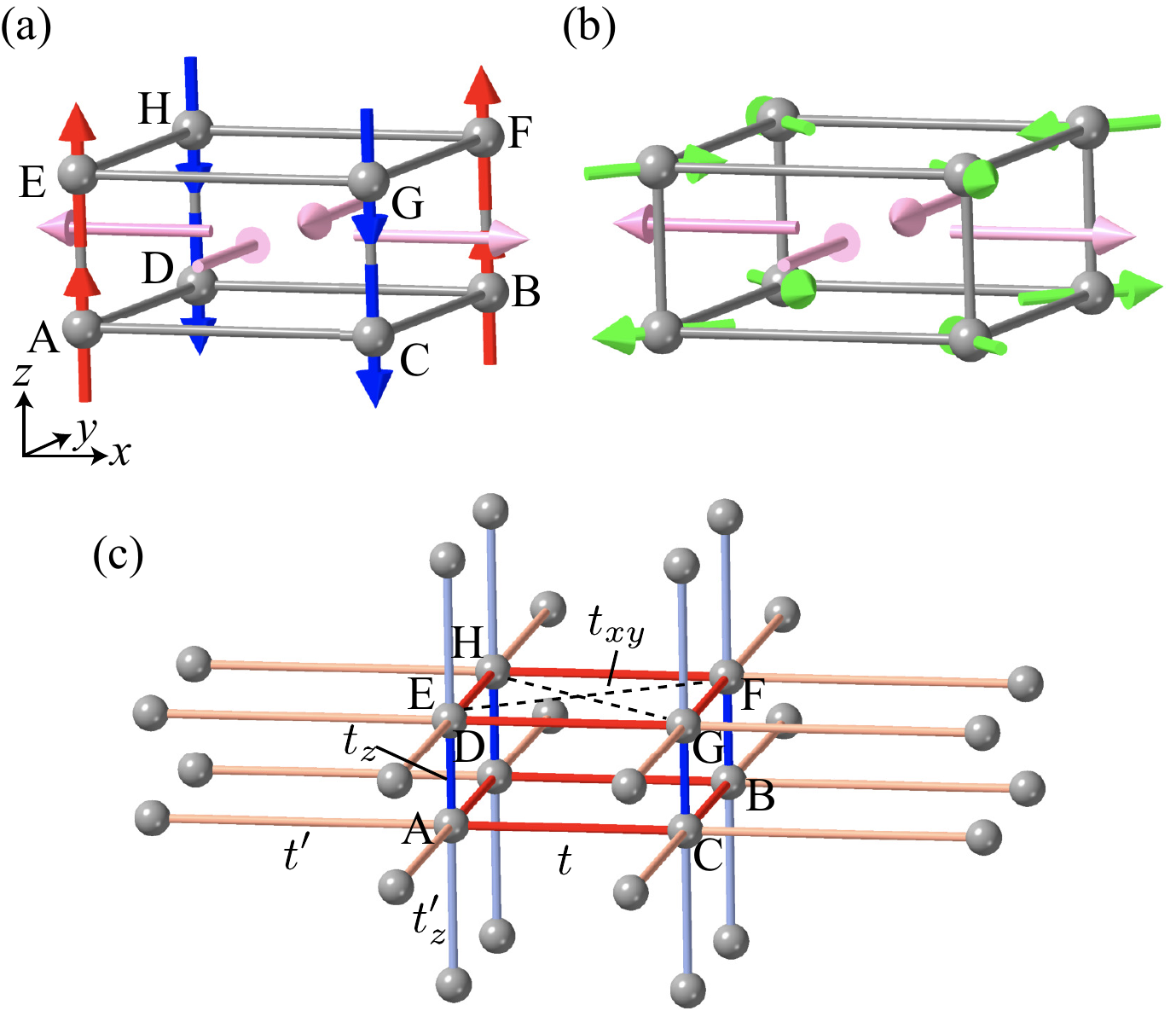} 
\caption{
\label{fig: fig2}
(Color online)
(a) Collienar and (b) noncollinear spin configurations with the MTQ $T_v$ in the rectangular solid consisting of the eight sublattices A--H. 
The red, blue, and green arrows on each sublattice represent spin moments with the positive, negative, and zero $S_i^z$, respectively, while the pink arrows in each face of the rectangular solid stand for the MTD. 
(c) The schematic picture of an eight-sublattice tetragonal system under the $4/mmm1'$ symmetry. 
}
\end{center}
\end{figure}

The expressions in Eqs.~(\ref{eq: Tu})--(\ref{eq: Tyz}) also give a relationship between the MTQ and the AFM spin configuration. 
To demonstrate that, we here consider an eight-sublattice rectangular solid, as shown in Fig.~\ref{fig: fig2}(a). 
When supposing that the basal plane is square, the eight-sublattice system belongs to the MPG $4/mmm1'$. 
By performing the multipole expansion for the magnetic cluster based on the virtual atomic cluster method~\cite{Suzuki_PhysRevB.99.174407}, one finds that five out of twenty-four AFM spin configurations possess nonzero MTQ moments and belong to the different irreducible representation from the magnetic dipole; the five irreducible representations are represented as $A^-_{1g}\oplus 2B^-_{1g}\oplus 2B^-_{2g}$ (the superscript stands for the time-reversal parity). 
Here, the irreducible representations of $A^{-}_{1g}$, $B^{-}_{1g}$, and $B^{-}_{2g}$ correspond to nonzero $T_u$, $T_v$, and $T_{xy}$, respectively. 

Among them, we examine two AFM orderings with $T_v$ as examples, which are characterized by the collinear and noncollinear spin configurations, as shown in Figs.~\ref{fig: fig2}(a) and \ref{fig: fig2}(b), respectively. 
In the noncollinar spin configuration in Fig.~\ref{fig: fig2}(b), each spin points along the $\langle 110 \rangle$ direction. 
In both AFM cases, the system reduces to $4'/mmm'$. 
Although one obtains nonzero $T_v$ for these AFM spin configurations by evaluating Eq.~(\ref{eq: Tv}), its appearance is intuitively understood from the spatial distribution of the MTD in each plaquette; the $T_v$-type distribution appears as shown by the pink arrows in Figs.~\ref{fig: fig2}(a) and \ref{fig: fig2}(b), which well corresponds to the distribution in Fig.~\ref{fig: fig1}(b).

Next, we consider the lattice system consisting of the eight-sublattice unit cell, as shown in Fig.~\ref{fig: fig2}(c). 
The $sd$ model Hamiltonian is given by 
\begin{align}
\mathcal{H}= - \sum_{ij \sigma}t_{ij} c_{i\sigma}^{\dagger}c_{j\sigma}^{} - \sum_{i\sigma \sigma'} \bm{h}_i \cdot c_{i\sigma}^{\dagger} \bm{\sigma}_{\sigma \sigma'}c_{i\sigma'}^{},
\label{eq:Ham}
\end{align}
where $c^{\dagger}_{i\sigma}$ ($c_{i\sigma}^{}$) is the creation (annihilation) operator for site $i$ and spin $\sigma=\uparrow, \downarrow$. 
The Hamiltonian consists of the hopping term with the five hopping parameters $(t, t', t_z, t'_z, t_{xy})$ in Fig.~\ref{fig: fig2}(c) and the AFM mean-field term to induce the spin configurations in Figs.~\ref{fig: fig2}(a) and \ref{fig: fig2}(b). 
For example, we set $\bm{h}_{\rm A}=(0,0,h)$ for the collinear spin configuration in Fig.~\ref{fig: fig2}(a) and $\bm{h}_{\rm A}=(-h,-h,0)$ for noncollinear one in Fig.~\ref{fig: fig2}(b).
In the following, we set $t=1$ as the energy unit and set $t'=0.5$, $t_z=0.6$, and $t'_z=0.3$. 
We take the equal lattice constants for both $x$ and $z$ directions for simplicity. 

We briefly discuss the stabilization mechanisms of the spin configurations, i.e., the origin of $\bm{h}_i$, in Figs.~\ref{fig: fig2}(a) and \ref{fig: fig2}(b). 
One of the mechanisms is the direct exchange interaction between the neighboring spins; the ferromagnetic (AFM) Heisenberg interaction along the $z$ ($x$ and $y$) directions favors the collinear spin configuration in Fig.~\ref{fig: fig2}(a), while the ferromagnetic (AFM) Heisenberg interaction along the $x$ and $y$ ($z$) directions in addition to the AFM interaction along the $\langle 110 \rangle$ direction can lead to the noncollinear spin configuration in Fig.~\ref{fig: fig2}(b)~\cite{comment_spinconfig}.  
Another mechanism is based on the effective magnetic interaction in itienrant magnets~\cite{hayami2021topological}; the instability toward the spin configurations in Figs.~\ref{fig: fig2}(a) and \ref{fig: fig2}(b) has been discussed in the double exchange model and the periodic Anderson model in the limit of the square ($t=t'$ and $t_z=t'_z=t_{xy}=0$)~\cite{Agterberg_PhysRevB.62.13816,hayami_PhysRevB.91.075104,Hayami_PhysRevB.90.060402} and cubic ($t=t'=t_z=t'_z$ and $t_{xy}=0$)~\cite{Alonso_PhysRevB.64.054408,hayami2014charge,Hayami_PhysRevB.89.085124,Hayami_PhysRevB.90.060402} lattices. 
In addition to these factors, although the SOC might play a role in determining the spin directions, we neglect it in order to examine the behavior driven by the magnetic phase transition.

\begin{figure}[t!]
\begin{center}
\includegraphics[width=1.0 \hsize ]{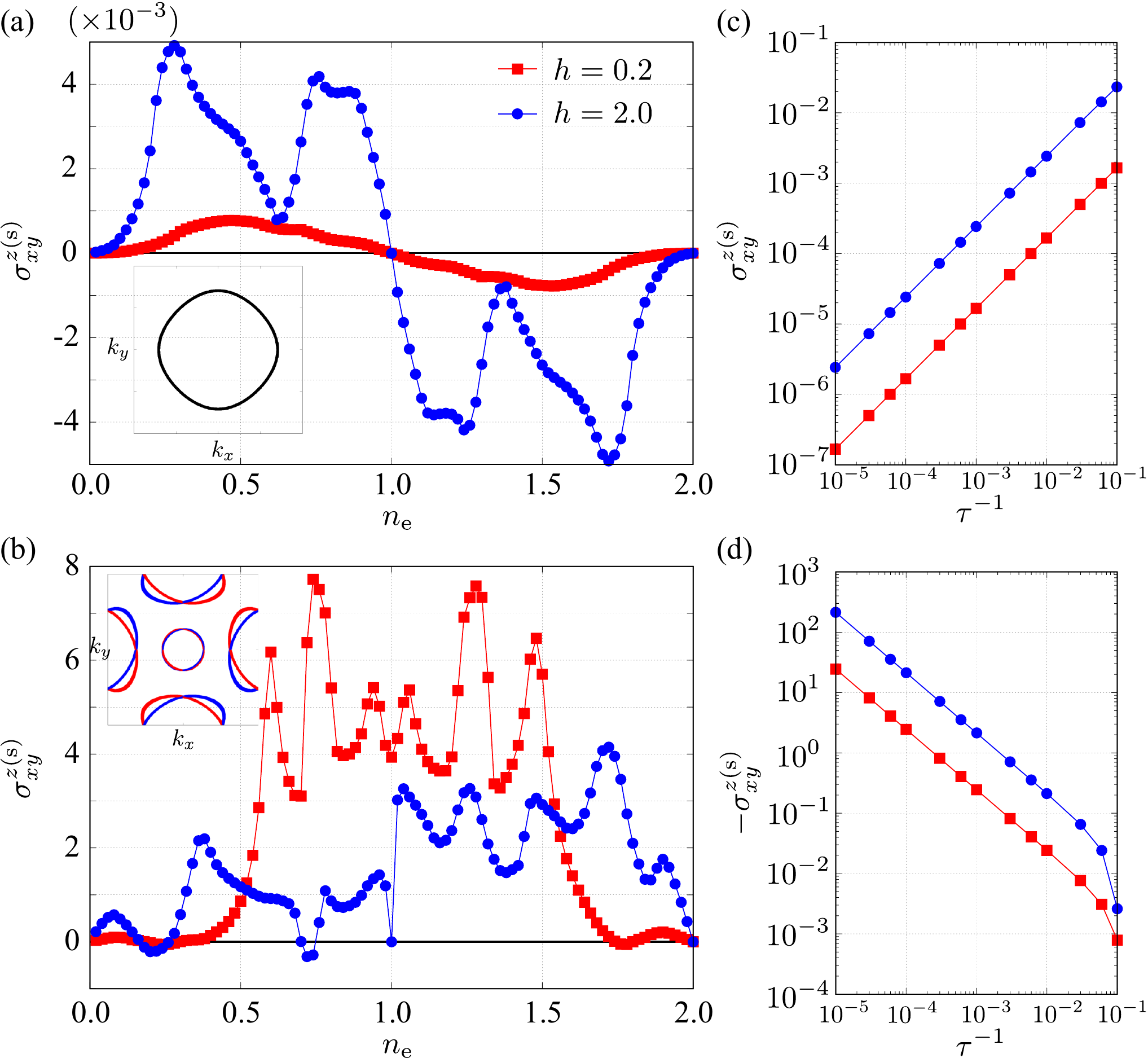} 
\caption{
\label{fig: fig3}
(Color online)
(a,b) $n_{\rm e}$ dependence of $\sigma^{z {\rm (s)}}_{xy}$ for $h=0.2$ and $h=2$ at (a) $t_{xy}=0$ and (b) $t_{xy}=0.2$ in the collinear AFM in Fig.~\ref{fig: fig2}(a). 
The inset of (a) [(b)] represents the isoenergy surfaces at $k_z=0$ and $\mu=-3.95$ ($\mu=-2.8$) in the first Brillouin zone, where $\mu$ represents the chemical potential; the red and blue colors in (b) stand for the up- and down-spin polarization, respectively.
(c,d) $\tau^{-1}$ dependence of $\sigma^{z {\rm (s)}}_{xy}$ with fixed $n_{\rm e}=0.2$ for $h=0.2$ and $h=2$ at (c) $t_{xy}=0$ and (d) $t_{xy}=0.2$. 
}
\end{center}
\end{figure}

As the MTQ is characterized by the rank-2 polar tensor, its emergence leads to various physical phenomena, such as the linear magneto-elastic effect and the nonlinear magnetoelectric effect~\cite{hayami2018microscopic, Yatsushiro_PhysRevB.104.054412}. 
Among them, we focus on the spin-conductivity tensor $\sigma^{\eta {\rm (s)}}_{\mu\nu}$ in $J^{\eta{\rm (s)}}_{\nu} = \sum_{\mu}\sigma^{\eta {\rm (s)}}_{\mu\nu} E_{\mu}$~\cite{Hayami_PhysRevB.98.165110}, which has often been refereed to as the magnetic spin Hall effect~\cite{Seemann_PhysRevB.92.155138,Zelezny_PhysRevLett.119.187204,zhang2018spin,vzelezny2018spin,kimata2019magnetic,Mook_PhysRevResearch.2.023065}; $J^{\eta{\rm (s)}}_{\nu}= J_{\nu} \sigma_\eta$ represents the spin current with the spin $\sigma_\eta$ and $E_\mu$ represents the electric field for $\mu,\nu,\eta=x,y,z$. 
We compute $\sigma^{\eta {\rm (s)}}_{\mu\nu}$ by evaluating the $J^{\eta{\rm (s)}}_{\nu}$-$J_{\mu}$ correlation function based on the Kubo formula following Ref.~\citen{Mook_PhysRevResearch.2.023065} with the scattering rate $\tau^{-1}=10^{-2}$ and the temperature $T=10^{-2}$, unless otherwise mentioned. 
The summation of the wave vector $\bm{k}$ is taken over $120^3$ grid points in the first Brillouin zone. 
Nonzero components of $\sigma^{\eta {\rm (s)}}_{\mu\nu}$ in the $4'/mmm'$ symmetry under the AFM orderings are given by $\sigma^{x {\rm (s)}}_{yz}$, $\sigma^{x {\rm (s)}}_{zy}$, 
$\sigma^{y {\rm (s)}}_{zx}$, $\sigma^{y {\rm (s)}}_{xz}$, $\sigma^{z {\rm (s)}}_{xy}$, and $\sigma^{z {\rm (s)}}_{yx}$. 

\begin{figure}[t!]
\begin{center}
\includegraphics[width=0.65 \hsize ]{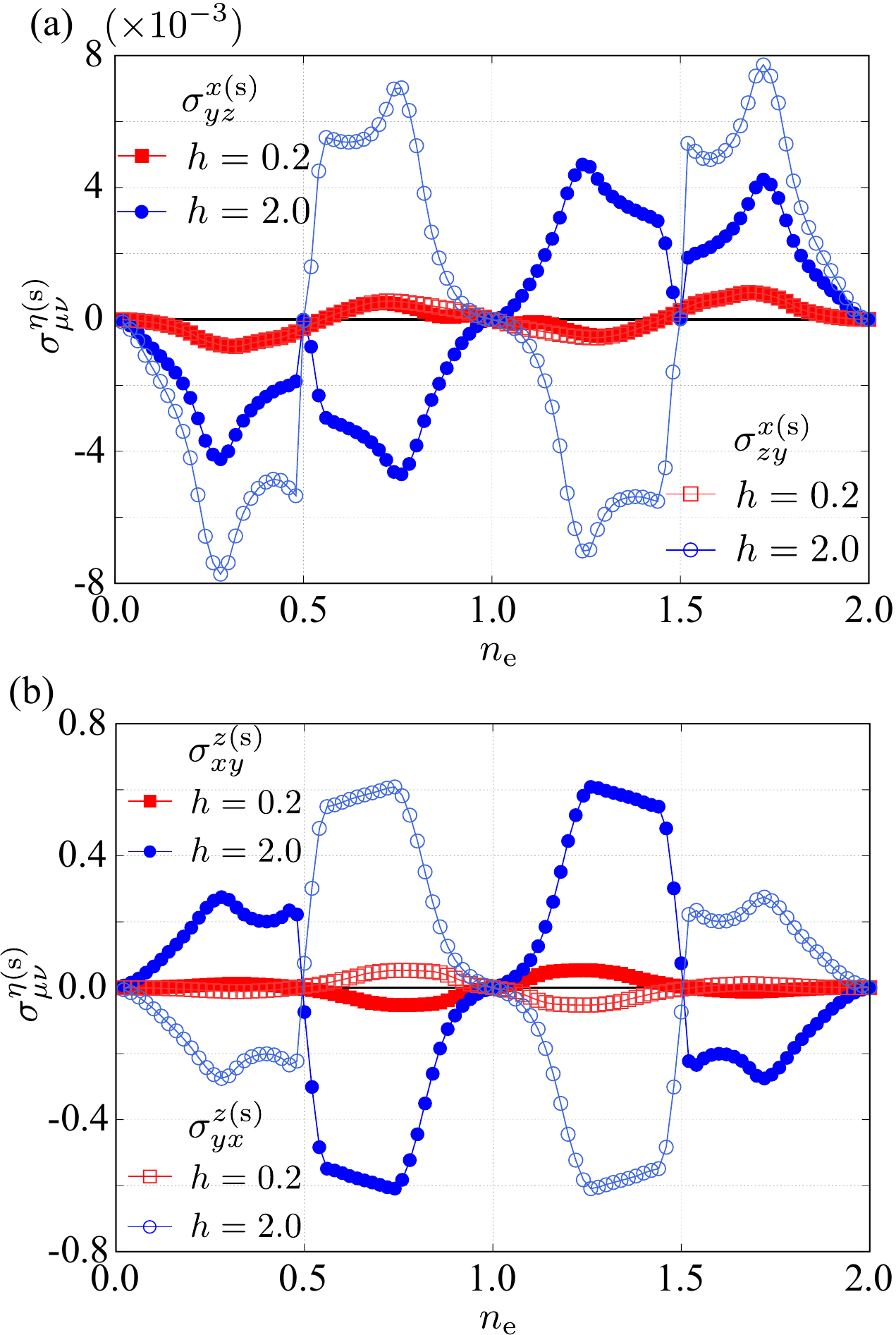} 
\caption{
\label{fig: fig4}
(Color online)
(a) $n_{\rm e}$ dependences of (a) $\sigma^{x {\rm (s)}}_{yz}$ and $\sigma^{x {\rm (s)}}_{zy}$ and (b) $\sigma^{z {\rm (s)}}_{xy}$ and $\sigma^{z {\rm (s)}}_{yx}$ for $h=0.2$ and $h=2$ at $t_{xy}=0$ in the noncollinear AFM in Fig.~\ref{fig: fig2}(b). 
}
\end{center}
\end{figure}

We first discuss the result for the collinear AFM in Fig.~\ref{fig: fig2}(a). 
Owing to the absence of the SOC, only the $z$ component of $\sigma^{\eta {\rm (s)}}_{\mu\nu}$ becomes nonzero, i.e., $\sigma^{x {\rm (s)}}_{yz} = \sigma^{x {\rm (s)}}_{zy}  = \sigma^{y {\rm (s)}}_{zx} = \sigma^{y {\rm (s)}}_{xz}=0$. 
Figures~\ref{fig: fig3}(a) and \ref{fig: fig3}(b) show the filling ($n_{\rm e}=\sum_{i\sigma}\langle c^{\dagger}_{i\sigma}c_{i\sigma} \rangle/8$) dependence of $\sigma^{z {\rm (s)}}_{xy}$ for $h=0.2$ and $h=2$ at (a) $t_{xy}=0$ and (b) $t_{xy}=0.2$. 
Both results in Figs.~\ref{fig: fig3}(a) and \ref{fig: fig3}(b) indicate that nonzero $\sigma^{z {\rm (s)}}_{xy}$ is obtained for small $h=0.2$ and large $h=2$ unless the system becomes insulating at  $n_{\rm e}=1$. 
Moreover, we confirm that only the symmetric component of the spin conductivity, i.e., $\sigma^{z {\rm (s)}}_{xy}=\sigma^{z {\rm (s)}}_{yx}$, appears, which is expected from the symmetry analysis in the presence of $T_v$~\cite{Hayami_PhysRevB.98.165110}.

Meanwhile, one finds that the magnitudes of $\sigma^{z {\rm (s)}}_{xy}$ in Figs.~\ref{fig: fig3}(a) and \ref{fig: fig3}(b) are substantially different from each other; the values with $t_{xy}=0.2$ is much larger than those with $t_{xy}=0$ by the order of $10^{3}$. 
Their difference is understood from the different mechanisms of $\sigma^{z {\rm (s)}}_{xy}$ that originates from the electronic band structures; the system with $t_{xy} \neq 0$ exhibits the spin-split band structure in the form of $k_x k_y \sigma_z$ [inset of Fig.~\ref{fig: fig3}(b)], while that with $t_{xy}= 0$ does not [inset of Fig.~\ref{fig: fig3}(a)]~\cite{naka2019spin,hayami2019momentum,Hayami2020b,Yuan_PhysRevB.102.014422}. 
The absence of the spin splitting with $t_{xy}= 0$ is attributed to the fact that there are no microscopic degrees of freedom in the hopping term to couple to the AFM mean fields~\cite{hayami2019momentum}.
Since the spin splitting as $k_x k_y \sigma_z$ indicates the direct coupling between the spin current $J_y \sigma_z \sim k_y \sigma_z$ ($J_x \sigma_z \sim k_x \sigma_z$) and input field $E_x$ ($E_y$) that flows the electric current in metals $J_x \sim k_x$ ($J_y\sim k_y$), $\sigma^{z {\rm (s)}}_{xy}$ is largely enhanced for $t_{xy}\neq 0$. 
In fact, the intraband process is dominant for $t_{xy}\neq 0$ [Fig.~\ref{fig: fig3}(b)], while only the interband one is present for $t_{xy}=0$ [Fig.~\ref{fig: fig3}(a)].
Such a difference is found in the $\tau^{-1}$ dependence of $\sigma^{z {\rm (s)}}_{xy}$ in Figs.~\ref{fig: fig3}(c) and \ref{fig: fig3}(d); the case for $t_{xy}=0$ ($t_{xy}\neq 0$) is proportional to $\tau^{-1}$ ($\tau$), which means that the interband (intraband) process is dominant.

To further examine the difference in Figs.~\ref{fig: fig3}(a) and \ref{fig: fig3}(b) from the viewpoint of the model-parameter dependence, we expand $\sigma^{z {\rm (s)}}_{xy}$ as a polynomial form of products of the Hamiltonian matrix at wave vector $\bm{k}$, $\mathcal{H}(\bm{k})$, and the velocity operator, $\bm{v}_{\bm{k}}=\partial \mathcal{H}(\bm{k})/\partial \bm{k}$, based on the procedure in Ref.~\citen{Oiwa_doi:10.7566/JPSJ.91.014701}. 
As a result, the lowest-order contribution to $\sigma^{z {\rm (s)}}_{xy}$ is given by $ht^2_{xy}$ for $t_{xy} \neq 0$, while that is given by $h (t^2-t'^2)^2$ for $t_{xy}=0$. 
This indicates that the complicated hopping path in real space is necessary in the case of $t_{xy}=0$, which tends to suppress $\sigma^{z {\rm (s)}}_{xy}$. 

Next, we discuss the spin conductivity for the noncollinear AFM in Fig.~\ref{fig: fig2}(b). 
In the noncollinear AFM, all the components allowed from the symmetry ($\sigma^{x {\rm (s)}}_{yz}$, $\sigma^{x {\rm (s)}}_{zy}$, 
$\sigma^{y {\rm (s)}}_{zx}$, $\sigma^{y {\rm (s)}}_{xz}$, 
$\sigma^{z {\rm (s)}}_{xy}$, and $\sigma^{z {\rm (s)}}_{yx}$) become nonzero. 
We show the behaviors of $\sigma^{x {\rm (s)}}_{yz}$ and $\sigma^{x {\rm (s)}}_{zy}$ in Fig.~\ref{fig: fig4}(a) and those of $\sigma^{z {\rm (s)}}_{xy}$ and $\sigma^{z {\rm (s)}}_{yx}$ in Fig.~\ref{fig: fig4}(b) against $n_{\rm e}$ for $h=0.2$ and $h=2$ at $t_{xy}=0$. 
We omit the results of $\sigma^{y {\rm (s)}}_{zx}$ and $\sigma^{y {\rm (s)}}_{xz}$, as they are related to $\sigma^{x {\rm (s)}}_{zy}$ and $\sigma^{x {\rm (s)}}_{yz}$ owing to the $4'$ symmetry. 
In contrast to the collinear AFM case, as shown in Fig.~\ref{fig: fig4}(a), there is an antisymmetric component between $\sigma^{x {\rm (s)}}_{zy}$ and $\sigma^{x {\rm (s)}}_{yz}$, i.e., $\sigma^{x {\rm (s)}}_{zy}-\sigma^{x {\rm (s)}}_{yz} \neq 0$, which is attributed to the noncollinear structure; $y$-spin component contributes to the difference between $\sigma^{x {\rm (s)}}_{zy}$ and $\sigma^{x {\rm (s)}}_{yz}$. 
Both $\sigma^{x {\rm (s)}}_{yz}$ and $\sigma^{x {\rm (s)}}_{zy}$ show similar behavior to that in Fig.~\ref{fig: fig3}(a), where only the interband process contributes to the spin conductivity; the diagonal hopping in the $xz$ and $yz$ plane like between sublattices A and H is necessary to enhance the spin conductivity through the intraband contribution owing to the spin-split band structure. 
For $\sigma^{x {\rm (s)}}_{yz}$ and $\sigma^{x {\rm (s)}}_{zy}$, the lowest-order essential model parameter dependence is given by $h (t^2-t'^2)(t_z^2-t'^2_z)$.

Another differece from the collinear AFM is found in $\sigma^{z {\rm (s)}}_{xy}$ and $\sigma^{z {\rm (s)}}_{yx}$, as shown in Figs.~\ref{fig: fig3}(a) and \ref{fig: fig4}(b). 
In the noncollinear AFM, the out-of-plane $z$-spin component of the spin conductivity also becomes nonzero, as pointed out in the previous studies~\cite{Zelezny_PhysRevLett.119.187204,zhang2018spin}. 
In the present noncollinear AFM structure, we obtain the antisymmetric spin conductivity to satisfy $\sigma^{z {\rm (s)}}_{xy}=-\sigma^{z {\rm (s)}}_{yx}$, as shown in Fig.~\ref{fig: fig4}(b). 
However, it is noted that the mechanism of $\sigma^{z {\rm (s)}}_{xy}$ and $\sigma^{z {\rm (s)}}_{yx}$ is different from that depending on $\tau^{-1}$ or $\tau$ in Figs.~\ref{fig: fig3}(c) and \ref{fig: fig3}(d); $\sigma^{z {\rm (s)}}_{xy}$ and $\sigma^{z {\rm (s)}}_{yx}$ does not show the $\tau$ dependence. 
In other words, the intrinsic interband process like the spin Hall effect in the nonmagnetic systems with the SOC~\cite{murakami2003dissipationless, Sinova_PhysRevLett.92.126603} and longitudinal spin conductivity in the systems with the electric toroidal dipole~\cite{hayami2021electric} is dominant for $\sigma^{z {\rm (s)}}_{xy}$ and $\sigma^{z {\rm (s)}}_{yx}$, where the vector chirality degree of freedom in the plaquette ACBD or EGFH plays a similar role to the SOC. 
Thus, this component is regarded as a secondary effect owing to the effective SOC under the noncollinear spin configuration rather than the MTQ-driven effect. 
The necessity of the noncollinear spin configuration is verified in the parameter expansion of $\sigma^{z {\rm (s)}}_{xy}$; the essential model parameters are proportional to $h^2$ as given by $h^2 (t^2-t'^2)^2$~\cite{comment_h2}. 

\begin{table}[t!]
\centering
\caption{
List of magnetic point groups (MPGs) to possess the MTQ as a primary order parameter in both centrosymmetric ($\mathcal{P}: \bigcirc$) and noncentrosymmetric ($\mathcal{P}: \times$) magnetic systems. 
Multipoles in the column ``Other" represent the MT monopole ($T_0$) and magnetic octupoles ($M_{xyz}$, $M_{3b}$, and $M_{z}^\beta$) that contributes to the spin conductivity tensor. 
The candidate materials are also shown in the rightmost column. 
 \label{table:mp}}
\begin{tabular}{p{2.4cm}p{0.3cm}p{0.7cm}p{1.0cm}p{2.2cm}}\hline 
&$\mathcal{P}$ &MTQ& Other & Material
\\ \hline
 
 $4/mmm,6/mmm$ & $ \bigcirc$ & $T_u$ & $ T_0$ &  CdYb$_2$(S,Se)$_4$~\cite{Dalmas_PhysRevB.96.134403}   \\

$422,\bar{4}2m, 4mm$ & $ \times$  & & & Ho$_2$Ge$_2$O$_7$~\cite{Morosan_PhysRevB.77.224423}
 \\
$622,\bar{6}m2,6mm$ & & & & CuFeS$_2$~\cite{Donnay_PhysRev.112.1917}
 \\
\hline

 $mmm$ &$ \bigcirc$ & $T_u, T_v$  & $  T_0, M_{xyz}$ & MnTe~\cite{kunitomi1964neutron}
 \\

$222,mm2$ &$ \times$ & & & ErGe$_{1.83}$~\cite{oleksyn1997structure}
\\
\hline

$\bar{3}m$  & $\bigcirc$ & $T_u$ & $ T_0, M_{3b}$ & CoF$_3$~\cite{lee2018weak} \\
 
$32,3m$ & $\times$ & & & Ba$_3$MnNb$_2$O$_9$~\cite{Lee_PhysRevB.90.224402}
 \\
 \hline
 
$4'/mmm'$ &$\bigcirc$ & $T_v$  
&   $M_{xyz}$ & CoF$_2$~\cite{jauch2004gamma} \\

$4'22',\bar{4}'2m'$  &$\times$ & & & Ce$_4$Sb$_3$~\cite{nirmala2009understanding}
 \\
$\bar{4}'m2',4'mm'$  &  & & & 
 \\
 \hline

$4'/m$ & $\bigcirc$ &
 $T_v$, $T_{xy}$  
&   $M_{xyz}$, $M_z^\beta$ \\

$4',\bar{4}'$ &$\times$ & & & CsCoF$_4$~\cite{lacorre1991ordered}
  \\

\hline 
\end{tabular}
\end{table}

So far, we have shown that the AFM with the MTQ in the $4'/mmm'$ system in Fig.~\ref{fig: fig2}(c) exhibits the characteristic spin conductivity. 
We discuss the possible magnetic systems from the symmetry viewpoint to stimulate experimental findings of the MTQ-related phenomena. 
Among all 122 types of MPGs, the MTQ becomes active for 42 MPGs~\cite{Yatsushiro_PhysRevB.104.054412}. 
Furthermore, for 22 out of 42 MPGs, the MTQ is regarded as a primary order parameter, as the lower-rank magnetic dipole is not activated. 
We list these 22 MPGs accompanying the MTQs in Table~\ref{table:mp}, where the information about the $\mathcal{P}$ symmetry, other activated multipoles contributing to the spin conductivity ($T_0$ represents the MT monopole and $M_{xyz}$, $M_{3b}$, and $M_{z}^\beta$ represent the magnetic octupoles)~\cite{Hayami_PhysRevB.98.165110, Yatsushiro_PhysRevB.104.054412}, and candidate materials reported in MAGNDATA~\cite{gallego2016magndata} are also shown. 
As these 22 MPG systems are not affected by the magnetic-dipole-related phenomena, one can expect pure MTQ-related phenomena. 

To summarize, we have investigated the MTQ, which corresponds to the time-reversal-odd rank-2 polar tensor degree of freedom, accompanied by the AFM spin configuration. 
By analyzing the $sd$ model in the presence of the AFM mean fields under the $4'/mmm'$ symmetry, we found that both collinear and noncollinear AFMs exhibit spin conductivity with the dissipation once the MTQ is activated. 
We have shown two types of mechanisms for spin conductivity: One arises from the interband process without the spin-split band structure, while the other arises from the intraband process induced by the spin-split band structure. 
We provided all the MPGs to possess the MTQ but without the magnetic dipole in order to stimulate exploration of MTQ-related physics.

\begin{acknowledgments}
This research was supported by JSPS KAKENHI Grants Numbers JP19K03752, JP19H01834, JP21H01037, and by JST PRESTO (JPMJPR20L8).
Parts of the numerical calculations were performed in the supercomputing systems in ISSP, the University of Tokyo.
\end{acknowledgments}

\bibliographystyle{JPSJ}
\bibliography{ref}

\begin{thebibliography}{10}
\bibitem{comment_spinconfig}
For the latter case, additional spin interactions, such as the ring-exchange
  and biquadratic interactions, are required to lift the degeneracy with the
  collinear-type spin configuration.
\bibitem{comment_h2}
When setting the different mean-field magnetidues for the $x$-spin and $y$-spin
  component as $h \to h_x$ and $h \to h_y$, respectively, the $h^2$ dependence
  is given by $h_x h_y$.
\end{thebibliography}

\end{document}